\documentclass[sigconf]{acmart}
\usepackage{tikz-cd}
\usepackage[utf8]{inputenc}
\usepackage[T1]{fontenc}
\usepackage[english]{babel}
\usepackage[small]{titlesec}
\usepackage{flushend}
\usepackage{etoolbox}
\makeatletter
\patchcmd{\ttlh@hang}{\parindent\z@}{\parindent\z@\leavevmode}{}{}
\patchcmd{\ttlh@hang}{\noindent}{}{}{}
\makeatother

\usepackage{todonotes}

\usepackage{balance} 

\usepackage{hyperref}
\hypersetup{
    colorlinks=true,
    linkcolor=blue,
    filecolor=red,
    urlcolor=magenta,
    breaklinks=true, 
}
\usepackage{breakurl} 

\usepackage{amsmath, listings, xspace, semantic, stmaryrd, mathrsfs, enumerate, mathpartir, mathtools}
\usepackage[threshold=1]{csquotes}

\usepackage{lstcontracts}
\usepackage{lstcoq}
\usepackage{lsthaskell}
\usepackage{lstfuthark}

\definecolor{ltblue}{rgb}{0,0.4,0.4}
\definecolor{dkblue}{rgb}{0,0.1,0.6}
\definecolor{dkgreen}{rgb}{0,0.35,0}
\definecolor{dkviolet}{rgb}{0.3,0,0.5}
\definecolor{dkred}{rgb}{0.5,0,0}

\newcommand\tuple[1]{(#1)}
\newcommand\synhor{\textsc{hor}}

\newcommand\ilSem[2]{{\mathcal {I\!L}}\left\llbracket#1\right\rrbracket_{#2}}
\newcommand\sem[1]{\left\llbracket#1\right\rrbracket}
\newcommand\tSem[2]{{\mathcal {T}}\left\llbracket#1\right\rrbracket_{#2}}
\newcommand\cSem[2]{{\mathcal C}\left\llbracket#1\right\rrbracket_{#2}}
\newcommand\gSem[1]{\left\llbracket#1\right\rrbracket}
\newcommand\tysem[1]{\left\llbracket#1\right\rrbracket}
\newcommand\compileExp[2] {{\tau_{\textrm{e}}}\left\llbracket#1\right\rrbracket_{#2}}
\newcommand\compileContr[2]{{\tau_{\textrm{c}}}\left\llbracket#1\right\rrbracket_{#2}}

\newcommand\eSem[2]{{\mathcal E}\left\llbracket#1\right\rrbracket_{#2}}
\newcommand\cRed[2]{\stackrel{#2}{\Longrightarrow_{#1}}}

\newcommand\smartTplus{\underline{\mathsf{tplus}}}
\newcommand\reals{{\mathbb R}}
\newcommand\nats{{\mathbb N}}
\newcommand\ints{{\mathbb Z}}
\newcommand\bools{{\mathbb B}}

\newcommand\type[1]{\mathtt{#1}}

\newcommand\acc{\mathtt{acc}}

\newcommand\TC[1]{\mathcal{T\!C}(#1)}

\newcommand\instdec{\mathsf{inst}}
\newcommand\inst[2]{\instdec(#1,#2)}
\newcommand\cutPayoff[1]{\mathsf{cutPayoff}(#1)}
\newcommand\zero{\mathtt{zero}}
\newcommand\scale[2]{\mathtt{scale}(#1,#2)}
\newcommand\both[2]{\mathtt{both}(#1,#2)}
\newcommand\transl[2]{\mathtt{translate}(#1,#2)}

\newcommand\transfer[3]{\mathtt{transfer}{(#1,#2,\mathtt{#3})}}

\newcommand\letA{\mathtt{let}}
\newcommand\letB{\mathtt{in}}

\newcommand\ifwithin[4]{\mathtt{ifWithin}(#1,#2,#3,#4)}
\newcommand\pto{\rightharpoonup}

\newcommand\id[1]{\ensuremath{\mathit{#1}}}

\reservestyle{\command}{\mathtt}
\command{if[if],model[model],not[not],neg[neg],loopif[loopif],
  payoff[payoff],add[add],sub[sub],mult[mult],div[div],lt[lt],and[and],true[true],
  or[or],scale[scale],translate[translate],zero[zero],
  transfer[transfer],both[both],obs[obs],now[now],let[let],Tnum[Tnum],Tvar[Tvar],
  ifWithin[ifWithin],rNum[rNum],bNum[bNum],Tplus[Tplus],Texpr[Texpr],TnumZ[TnumZ],cond[cond],tplus[tplus]}

\newcommand\semarg{\overline{\mathcal{P}}}

\newcommand{\icode}[1]{\lstinline[mathescape]!#1!}

\newtheorem{theorem}{Theorem}

\newtheorem{lemma}{Lemma}

\theoremstyle{remark}
\newtheorem{remark}{Remark}

\theoremstyle{definition}
\newtheorem{defn}{Definition}

\title[Certified Compilation of Financial Contracts]{Certified Compilation of Financial Contracts}

\author{Danil Annenkov}
\affiliation{INRIA\\GALLINETTE Research team}
\email{danil.annenkov@inria.fr}

\author{Martin Elsman}
\affiliation{University of Copenhagen\\
    Dept. of Computer Science (DIKU)
}
\email{mael@di.ku.dk}


\begin{document}

\begin{abstract}
We present an extension to a certified financial contract management
system that allows for templated declarative financial contracts
and for integration with financial stochastic models through verified compilation
into so-called payoff-expressions. Such expressions readily allow for
determining the value of a contract in a given evaluation context,
such as contexts created for stochastic simulations.
The templating mechanism is useful both at the contract specification
level, for writing generic reusable contracts, and for reuse of code
that, without the templating mechanism, needs to be recompiled for
different evaluation contexts.
We report on the effect of using the certified system in the context
of a GPGPU-based Monte Carlo simulation engine for pricing various
over-the-counter (OTC) financial contracts. The full
contract-management system, including the payoff-language compilation,
is verified in the Coq proof assistant and certified Haskell code is
extracted from our Coq development along with Futhark code for use in
a data-parallel pricing engine.
\end{abstract}

\copyrightyear{2018}
\acmYear{2018}
\setcopyright{acmlicensed}
\acmConference[PPDP '18]{The 20th International Symposium on Principles and Practice of Declarative Programming}{September 3--5, 2018}{Frankfurt am Main, Germany}
\acmBooktitle{The 20th International Symposium on Principles and Practice of Declarative Programming (PPDP '18), September 3--5, 2018, Frankfurt am Main, Germany}
\acmPrice{15.00}
\acmDOI{10.1145/3236950.3236955}
\acmISBN{978-1-4503-6441-6/18/09}


\maketitle

\section{Background and Motivation}
\lstset{language=Contracts}
New technologies are emerging that have potential for seriously disrupting the
financial sector. In particular, blockchain technologies, such as Bitcoins
\cite{nakamoto2008bitcoin} and the Ethereum Smart Contract peer-to-peer platform
\cite{ethereumyellow2015}, have entered the realm of the global financial market
and it becomes essential to ask to which degree users can trust that the
underlying implementations are really behaving according to the specified
properties. Unfortunately, the answers are not clear and errors may result in
irreversible high-impact events.

Contract description languages and payoff languages are used in large scale
financial applications \cite{MLFi, SimCorpXpress}, although formalisation of
such languages in proof assistants and certified compilation schemes are much
less explored.

The work presented here builds on a series of previous work on specifying
financial contracts
\cite{andersen06sttt,Arnold95analgebraic,Frankau09JFP,hvitved11jlap,SPJ2000} and
in particular on a certified financial contract management engine and its
associated contract DSL \cite{BahrBertholdElsman}. This framework allows for
expressing a wide variety of financial contracts (a fundamental notion in
financial software) and for reasoning about their functional properties (e.g.,
horizon and causality).
As in the previous work, the contract DSL that we consider is equipped with a
denotational semantics, which is independent of stochastic aspects and depends
only on an \emph{external environment}
$\type{ExtEnv} :\mathbb{N} \times \type{Label} -> \mathbb{R} + \mathbb{B}$,
which maps observables (e.g., the price of a stock on a particular day) to
values.
%
%
As the first contribution of this work, we present a certified compilation
scheme that compiles a contract into a \emph{payoff function}, which
aggregates all cashflows in the contract, after discounting them according to
some model. The result represents a single ``snapshot'' value of the
contract. The payoff language is inspired by traditional payoff languages and is
well suited for integration with Monte Carlo simulation techniques for
pricing. It is essentially a small expression language featuring arithmetic and
boolean operators, a limited form of a looping construct, and enriched with
notation for looking up observables in the external environment. We show that
compilation from the contract DSL to the payoff language preserves the cashflow
semantics.

The contract language described in \cite{BahrBertholdElsman}, deals
with concrete contracts, such as a one year European call
option on the AAPL (Apple) stock with strike price \$100. The lack of
genericity means that each time a new contract is created (even a
very similar one), the contract management engine needs to compile the
contract into the payoff language and further into a target language
for embedding into the pricing engine. As our second contribution,
we introduce the notion of a \emph{financial instrument},
which allows for templating of contracts and which can be turned into
a concrete contract by instantiating template variables with
particular values. For example, an European call option instrument has
template parameters such as maturity (the end date of the contract),
strike, and the underlying asset that the option is based
on. Compiling such a template once allows the engine to reuse compiled
code, giving various parameter values as input to the pricing engine.

Moreover, an inherent property of contracts is that they evolve over time. This
property is precisely captured by a contract reduction semantics. Each day, a
contract becomes a new ``smaller'' contract, thus, for pricing purposes,
contracts need to be recompiled at each time step, resulting in a dramatic
compilation time overhead. As a third contribution, we introduce a mechanism
allowing for avoiding recompilation in relation to contract evolution. A
payoff expression can be parameterised over the \emph{current time} so that
evaluating the payoff code at time $t$ gives us the same result (up to
discounting) as first advancing the contract to time $t$, then compiling it to
the payoff code, and then evaluating the result. Most of the payoff languages
used in real-world applications require synchronization of the contract and the
payoff code once a contract evolves \cite[Contract State and Pricing
  Synchronization]{MLFi-whitepaper}. In some cases, however, as we mentioned
earlier, it is important to capture the reduction semantics in the payoff
language as well. Our result allows for using a single compilation procedure
for both use cases: compiling a contract upfront and synchronizing at each time
step.

The contract analyses and the contract transformation procedures form a core code
base, which financial software crucially depends on. A certified
programming approach using the Coq proof assistant allows us to prove
the above desirable properties and to extract certified executable
code.

We summarize the contributions of this paper as follows:
\begin{itemize}
  \item We present an extended domain-specific language for expressing financial
    contracts, called CL, based on the work by \cite{BahrBertholdElsman}.  The
    extended contract language features contract templates, also called
    \emph{instruments}. The extension allows for parameterisation of contracts
    with respect to temporal parameters.
\item Inspired by traditional payoff languages, we develop a payoff
  intermediate language, which we demonstrate is well-suited for the integration with
  Monte Carlo simulation techniques.
\item We use the Coq proof assistant to develop a certified compilation
  procedure of contract templates into a parameterised payoff intermediate language.
\item We further parameterise the compiled payoff expressions with the notion of ``current time''
  allowing for capturing the evolution of contracts with the passage of time.
\item We develop the proof of an extended soundness theorem in the Coq proof
  assistant. The theorem establishes a correspondence between the
  time-parameterised compilation scheme and the contract reduction semantics.
\item We demonstrate how the parametric payoff code allows for better performance due to
  avoiding recompilation with the change of parameters.
\end{itemize}

\section{The Contract Language}\label{subsec:contracts:contracts-syn-sem}

The contract language that we consider follows the style of
\cite{BahrBertholdElsman} and is a declarative contract language that
allows for expressing contractual agreements among parties regarding
immediate and future cash flows between the parties. Contracts may
refer to observable values and cash flows may thereby depend on the
development of such observables, which may include stock prices,
interest rates, and decisions made by parties.

We assume a countably infinite set of program variables, ranged over by
$\id{v}$. Moreover, we use $\id{n}$, $\id{i}$, $\id{f}$, and $\id{b}$ to range
over natural numbers, integers, floating point numbers, and booleans. We use $p$
to range over parties and $a$ to range over assert symbols (e.g., EUR, USD, and
so on).
The contract language is given in Figure~\ref{fig:contracts:syntax};
it follows closely the contract language of \cite{BahrBertholdElsman},
but is extended with template variables.
\begin{figure}
\begin{eqnarray*}
  c \in \type{Contr} & ::= &  \<zero>~|~\<transfer>(p_1,p_2,a)~|~\<scale>(e,c)\\
  & ~|~ & \<translate>(t,c) ~|~ \<ifWithin>\left(e,t,c_1,c_2\right) \\
  & ~|~ & \<both>(c,c)\\
  e \in \type{Exp}  &::=&  op(e_1,~e_2, ~\dots,~e_n) ~|~ \<obs>(l,i)~|~ f ~|~ b\\
  t \in \type{TExp} & ::=&  n ~|~ v \\
  op \in \type{Op} &::=& \<add> ~|~ \<sub> ~|~ \<mult> ~|~ \<lt> ~|~ \<neg> ~|~ \<cond> ~|~\dots
\end{eqnarray*}
\caption{Syntax of contracts, contracts expressions, and template expressions.}\label{fig:contracts:syntax}
\end{figure}
\noindent
Expressions ($e \in \type{Exp}$) may contain \emph{observables}, which are
interpreted in an external environment. A contract may be empty ($\<zero>$), a
transfer of one unit (for simplicity) ($\<transfer>$), a scaled contract
($\<scale>$), a translation of a contract into the future ($\<translate>$), the
composition of two contracts ($\<both>$), or a generalized conditional
$\<ifWithin>(cond,t,c_1,c_2)$, which checks the condition $\id{cond}$ repeatedly
during the period given by $t$ and evaluates to $c_1$ if $\id{cond} = \<true>$
or to $c_2$ if $\id{cond}$ never evaluates to $\<true>$ during the period $t$.

The main difference between the original version of the contract language
and the version presented here is the introduction of \emph{template
  expressions} ($\id{t}$), which, for instance, allows us to write contract
templates with the contract maturity as a parameter.  This feature requires
refined reasoning about the temporal properties of contracts, such as
causality. Certain constructs in the original contract language, such as
$\<translate>(n,c)$ and $\<ifWithin>(\id{cond},n,c_1,c_2)$, are designed such
that basic properties of the contract language, including the property of
causality, are straightforward to reason about. In particular, the displacement
numbers $n$ in the above constructs are constant positive numbers. For
templating, we refine the constructs to support template expressions in place of
positive constants.
One of the consequences of adding template variables is that the
semantics of contracts now depends also on mappings of template
variables in a \emph{template environment} $\type{TEnv}: \type{Var} ->
\mathbb{N}$, which is also the case for many temporal properties of
contracts. For example, the type system for ensuring causality of
contracts \cite{BahrBertholdElsman} and the concept of horizon are now
parameterized by template environments.

Let us consider a few examples of contracts written in English and expressed
in CL.
\begin{example}\label{ex:contracts:option}
A \emph{European option} is a contract that gives the owner the right,
but not the obligation, to buy or sell an underlying security at a
specific price, known as the strike price, on the option's expiration
date (investopedia.com).

Let us take the expiration date to be 90 days
into the future and set the strike at USD 100.  In CL, we can
implement the European option contract with the above parameters as
follows:
\begin{lstlisting}
   translate(90,
     if(obs(AAPL,0) > 100.0,
        scale(obs(AAPL,0) - 100.0, transfer(you, me, USD)),
        zero))
\end{lstlisting}
\end{example}

\begin{example}
  A Three month FX swap for which the payment schedule has been
  settled is easily expressed in CL:
\begin{lstlisting}
   scale(1.000.000,
     both(
        all[translate(22, transfer(me, you, EUR)),
            translate(52, transfer(me, you, EUR)),
            translate(83, transfer(me, you, EUR))],
        scale(7.21,
         all[translate(22, transfer(you, me, DKK)),
             translate(52, transfer(you, me, DKK)),
             translate(83, transfer(you, me, DKK))])))
\end{lstlisting}

\noindent
In the swap-example, we have written
$\texttt{all[}c_1,\cdots,c_n\texttt{]}$ as an abbreviation for the
contract $\texttt{both(}c_1,\texttt{both(}\cdots,c_n\texttt{)}\texttt{)}$.
We use the $\texttt{all}$ shortcut with the \icode{translate} combinator to
implement a schedule of payments.
\end{example}

Using the CL template extension, we can abstract some of the contract
parameters in Example~\ref{ex:contracts:option} to template variables
(T for expiration date, and S for strike)\footnote{In our
  implementation, we focus on contract templates allowing for template
  expressions to represent temporal parameters, such as
  maturity. Other parameters, such as a strike, can be expressed as
  constant observable values.}:
\begin{lstlisting}
   translate(T,
     if(obs(AAPL,0) > S,
        scale(obs(AAPL,0) - S, transfer(you, me, USD)),
        zero))
\end{lstlisting}

This possibility for parameterisation plays well with how users would
interact with a contract management system. Contract templates could
be exposed to users as so-called \emph{instruments}, which a user can
instantiate to contracts by supplying concrete values for parameters.

We extend the denotational semantics from \cite{BahrBertholdElsman} to
accommodate the idea of template expressions. The semantics for the expression
sublanguage stays unchanged, since such expressions do not contain template
expressions. That is, the semantics for an expression $e \in \type{Exp}$ in
Figure~\ref{fig:contracts:syntax} is given by the partial function $\eSem{e}{} :
\gSem\Gamma \times \type{Env} \pto \tysem\tau$.  On the other hand, we modify
the semantic function for contacts by adding a template environment as an
argument:
\begin{align*}
  \cSem{c}{} &: \gSem{\Gamma} \times \type{Env} \times \type{TEnv} \pto \type{Trace}\\
  \type{Trace} &= \mathbb{N} \to \type{Trans}\\
  \type{Trans} &= \type{Party}\times\type{Party}\times\type{Asset} \to \mathbb{R}
\end{align*}
As the original contract semantics, it depends on the external environment
$\type{Env}:\mathbb{N} \times \type{Label} -> \mathbb{R} \cup \mathbb{B}$ and variable
assignments that map each free variable of type $\tau$ to a value in
$\tysem\tau$ with $\tysem{\type{Real}} = \reals$ and $\tysem{\type{Bool}} = \bools$.
Given a typing environment $\Gamma$, the set of \emph{variable assignments} in
$\Gamma$, written $\gSem\Gamma$, is the set of all partial mappings $\gamma$
from variable names to $\reals\cup\bools$ such that $\gamma(x)\in\tysem\tau$ iff
$x : \tau \in \Gamma$. The typing rules also remain the same for expressions and
for contracts.

The semantics for template expressions $\tSem{t}{} : \type{TEnv} \to \nats$ is defined as follows:
\[ \tSem{n}\delta = n \qquad \qquad \tSem{v}\delta = \delta(v) \]

We modify the semantics of contract constructors that depend on
template expressions in such a way that the corresponding template
expression is evaluated using $\tSem{-}{}$.  For example for
the $\<translate>$ constructor, we have
\[ \cSem{\transl t c}{\gamma,\rho,\delta} =
   \id{delay}(\tSem{T}{\delta},\cSem{c}{\gamma,\rho,\delta}) \]
\noindent
Where $\mathit{delay} : \nats \times \type{Trace} -> \type{Trace}$ is an
operation that delays a given trace by a number of time steps (see \cite[Figure~4]{BahrBertholdElsman}).

We define an \emph{instantiation function} that takes a contract and a template environment
containing values for template variables, and produces another contract that does not
contain template variables by replacing all occurrences of template variables with corresponding
values from the template environment.
\begin{defn}[Instantiation function]\label{eq:contratcs:inst-func}
\begin{multline*}
\begin{aligned}
  \instdec &: \type{Contr} \times \type{TEnv} -> \type{Contr}\\
  \inst {\zero} {\delta} &= \zero\\
  \inst {\letA~e~\letB~c} {\delta} &= \inst c \delta\\
  \inst {\transfer{p_1}{p_2}a} \delta &= {\transfer{p_1}{p_2}{a}}\\
  \inst {\scale e c} \delta &= \inst c \delta\\
  \inst {\transl t c} \delta &= \transl{\tSem{t}{\delta}}{\inst{c}{\delta}}\\
  \inst {\both{c_1}{c_2}} \delta &= \both{\inst{c_1}{\delta}}{\inst{c_2}{\delta}}
\end{aligned}\\
  \inst {\ifwithin e t {c_1} {c_2}} \delta = \\
  \ifwithin{e}{\tSem{t}{\delta}}{\inst{c_1}\delta} {\inst{c_2}\delta}
\end{multline*}
\end{defn}

We further define an inductive predicate that holds only for contract expressions
without template variables (Figure~\ref{fig:contracts:template-closed}). We call such
contracts \emph{template-closed}.
\begin{figure}
  \fbox{$\TC c$}
  \begin{mathpar}
  \inferrule{ }{\TC \zero}
  \and
  \inferrule{\TC c}{\TC {\letA~e~\letB~c}}
  \and
  \inferrule{ }{\TC{\transfer{p_1}{p_2}{a}}}
  \and
  \inferrule{\TC c}{\TC {\scale e c}}
  \and
  \inferrule{n~\text{is a numeral} \\ \TC c}{{\transl n c}}
  \and
  \inferrule{\TC{c_1} \\ \TC{c_2}}{\TC {\both{c_1}{c_2}}}
  \and
  \inferrule{n~\text{is a numeral} \\ \TC{c_1} \\ \TC{c_2}}
            {\TC {\ifwithin e n {c_1} {c_2}}}
\end{mathpar}
  \caption{Template-closed contracts.}\label{fig:contracts:template-closed}
\end{figure}

It is straightforward to establish the following fact.
\begin{lemma}
  For any contract $c$ and template environment $\delta$, application
  of the instantiation function gives a template-closed contract:
  \[\TC{\inst c \delta}\]
\end{lemma}

\begin{lemma}[Instantiation soundness]
  For any contract $c$, template environments $\delta$ and $\delta'$, external environment
  $\rho$, and any value environment $\gamma$, the contract $c$ and
  $\inst c \delta$ are semantically equivalent.
  That is, $\cSem{c}{\gamma,\rho,\delta} = \cSem{\inst{c}{\delta}}{\gamma,\rho,\delta'}$.
\end{lemma}

The reduction semantics of the contract language presented in \cite{BahrBertholdElsman}
remains the same, although, we make additional assumption that the contract expression is
closed with respect to template variables.


\section{The Payoff Intermediate Language}
The contract language allows for capturing different aspects of financial
contracts.  We consider a particular use case for the contract language, where
one wants to calculate an estimated price of a contract according to some
stochastic model by performing simulations. Simulations is often implemented
using Monte Carlo techniques, for instance, by evaluating a contract price at
current time for randomly generated possible market scenarios and discounting
the outcome according to some model. A software component that implements such a procedure
is called a \emph{pricing engine} and aims to be very efficient in performing
large amount of calculations by exploiting the parallelism
\cite{Andreetta:2016:FPF:2952301.2898354}. For this use case, one has to
 take the following aspects into account:
\begin{itemize}
\item Contracts should be represented as simple functions that take prices of
  assets involved in the contract (randomly generated by a pricing engine) and
  return one value corresponding to the aggregated outcome of the contract.
\item The resulting value of the contract should be discounted according to a
  given discount function.
\end{itemize}
One way of achieving this would be to implement an interpreter for the contract
language as part of a pricing engine. Although this approach is quite general,
interpreting a contract in the process of pricing will cause significant
performance overhead.  Moreover, it will be harder to reason about correctness
of the interpreter, since it could require non-trivial encoding in languages
targeting GPGPU devices. For that reason we take another approach: translating a
contract from CL to an intermediate representation and, eventually, to a
function in the pricing engine implementation language. We call this
intermediate representation a \emph{payoff language} and expressions in this
language we call \emph{payoff expressions}.

We would like the payoff language to contain fewer domain-specific
features and being closer to a subset of some general purpose
language, making a mapping from the payoff language to a target
language straightforward. We demonstrate how payoff expressions can be
translated to Haskell and Futhark
\cite{henriksen2014size,Henriksen:Futhark} in
Section~\ref{subsec:contracts:codegen}.
\begin{eqnarray*}
  \id{il} & ::= & \<now> ~|~\<model>(l,t) ~|~ \<if>(\id{il},\id{il},\id{il}) ~|~ \<loopif>(\id{il},\id{il},\id{il},t) \\ & ~|~ & \<payoff>(t,p,p) ~|~ \id{unop}(\id{il}) ~|~ \id{binop}(\id{il},\id{il})\\
  \id{unop} & ::= & \<neg> ~|~ \<not> \\
  \id{binop} & ::= &\<add>~|~\<mult>~|~\<sub>~|~\<lt>~|~\<and> ~|~\<or> ~|~ \dots\\
   t & ::= & n ~|~ i ~|~ v ~|~ \<tplus>(t,t)
\end{eqnarray*}

The payoff language is an expression language ($\id{il}\in\type{ILExpr}$) with
binary and unary operations, extended with conditionals and generalized
conditionals $\<loopif>$, behaving similarly to $\<ifWithin>$.  Template
expressions ($t\in\type{TExprZ}$) in this language are extensions of the
template expressions of the contract language with integer literals and
addition.

The semantics of payoff expressions is given in
Figure~\ref{fig:contracts:payoffs-sem}.  We use the notation $\semarg
= (\rho,\delta,t_0,t,d,p_1,p_2)$ for the vector of arguments to the
semantic function. The following notation $\semarg[t_0:=t_0'] =
(\rho,\delta,t_0',t,d,p_1,p_2)$ is used to show that the respective
argument in the vector receives a certain value. The semantics depends
on environments $\rho \in \type{Env}$ and $\delta \in \type{TEnv}$
similarly to the semantics of the contract language. Payoff
expressions can evaluate to a value of type $\nats$, $\reals$, or
$\bools$ (in contrast to the contract language for which the semantics
is given in terms of traces.).  We add $\nats$ to the semantic domain,
because we need to interpret the $\<now>$ construct, which represents
the ``current time'' parameter and template expressions $t^e$.  The
semantics also depends on a \emph{discount function} $d : \mathbb{N}
-> \mathbb{R}$.  The $t_0 \in \mathbb{N}$ parameter is used to add
relative time shifts introduced by the semantics of $\<loopif>$; $t$
is a current time, which will be important later, when we introduce a
mechanism to cut payoffs before a certain point in time.

The semantics for unary and binary operations is a straightforward mapping to
corresponding arithmetic and logical operations, provided that the arguments
have appropriate types. For example, $\sem{\<add>}(v_1, v_2) = v_1 + v_2$, if
$v_1, v_2 \in \reals$.

The semantic function $\ilSem{-}{}$ considers only payoffs between two parties
$p_1$ and $p_2$, which are given as the last two parameters. More precisely, it
considers payoffs from party $p_1$ to party $p_2$ as positive and as negative,
if payoffs go in the opposite direction. Another way of defining the semantics
could be a \emph{bilateral view} on payoffs. In this case only cashflows to or
from one fixed party to any other party are considered. Then, the semantics for
the $\<payoff>$ construct would be defined as follows:

\begin{gather*}
    \begin{aligned}[t]
      \ilSem{\<payoff>(t,p_1',p_2')}{\rho,\delta,t_0,t,d,p} &=
      \begin{cases}
        d (\tSem{t}{\delta}) &\text{if }  p_2' = p\\
        - d (\tSem{t}{\delta}) &\text{if } p_1' = p\\
        0 &\text{otherwise}
      \end{cases}
    \end{aligned}
\end{gather*}

\begin{figure}
{\footnotesize
\[
\begin{aligned}
  \boxed{\ilSem{il}{} :
  \type{Env} \times \type{TEnv} \times \mathbb{N} \times \mathbb{N} \times
  (\mathbb{N} -> \mathbb{R}) \times \type{Party} \times \type{Party}
  \pto \nats \cup \reals \cup \bools}
\end{aligned}
\]
}
\[
\begin{array}{rl}
  \semarg &= (\rho,\delta,t_0,t,d,p_1,p_2)
\end{array}
\]
\[
\begin{split}
\def\arraystretch{2}
\begin{array}{rl}
  \ilSem{t^e}{\semarg} &= \tSem{t^e}{\delta} + t_0 \\
  \ilSem{unop(il)}{\semarg} &=
  \sem{unop} (\ilSem{il}{\semarg})\\
  \ilSem{binop(il_0,il_1)}{\semarg} &=
  \sem{binop} (\ilSem{il_0}{\semarg},~\ilSem{il_1}{\semarg})
\end{array}
\end{split}
\]
\[
\begin{split}
\def\arraystretch{2}
\begin{array}{rl}
  \ilSem{\<model>(l,t^e)}{\semarg} &=
  \rho(l,\tSem{t^e}{\delta} + t_0)\\
  \ilSem{\<now>}{\semarg} &= t\\
  \ilSem{if(il_0,il_1,il_2)}{\semarg} &=
      \begin{cases}
        \ilSem{il_1}{\semarg},~
        \text{if } \ilSem{il_0}{\semarg} = true \\
        \ilSem{il_2,}{\semarg},~
        \text{if } \ilSem{il_0}{\semarg} = false \\
      \end{cases}\\[1em]
\end{array}
\end{split}
\]
\begin{multline*}
    \ilSem{\<payoff>(t^e,p_1',p_2')}{\semarg} =\\
      \begin{cases}
        d (\tSem{t^e}{\delta}) &\text{if }  p_1' = p_1, p_2' = p_2\\
        - d (\tSem{t^e}{\delta}) &\text{if } p_1' = p_2, p_2' = p_1\\
        0 &\text{otherwise}
      \end{cases}\\
  \ilSem{\<loopif>(il_0,il_1,il_2, t^e)}{\semarg} =
      iter(\tSem{t^e}{\delta},t_0), where\\[1em]
  iter(n,t_0') =
  \begin{cases}
    \ilSem{il_1}{\semarg[t_0:=t_0']},\\
    \quad \quad \text{if } \ilSem{il_0}{\semarg[t_0:=t_0']} = true\\
    \ilSem{il_2}{\semarg[t_0:=t_0']},~\\
    \quad \quad \text{if } \ilSem{il_0}{\semarg[t_0:=t_0']} = false \wedge n = 0 \\
    iter (n-1) (t'+1),~\\
    \quad \quad \text{if } \ilSem{il_0}{\semarg[t_0:=t_0']} = false \wedge i > 0
  \end{cases}
\end{multline*}
  \caption{Semantics of payoff expressions.}
  \label{fig:contracts:payoffs-sem}
\end{figure}

\section{Compiling Contracts to Payoffs}\label{subsec:contracts:payoffs-syn-sem}
The contract language consist of two levels, namely constructors to build
contracts ($c \in \type{Contr}$) and expressions used in some of these constructors ($\<scale>$,
$\<ifWithin>$, etc.).  We compile both levels into a single payoff language. The
compilation functions $\compileExp{-}{} : \type{Expr} \times \type{TExprZ} \pto
\type{ILExpr}$ and $\compileContr{-}{} : \type{Contr} \times \type{TExprZ} \pto
\type{ILExpr}$ are recursively defined on the syntax of expressions and
contracts, respectively, taking the starting time $t_0 \in \type{TExprZ}$ as a
parameter.
\[
\begin{aligned}[t]
      \compileExp{\<cond>(b,~e_0,~e_1])}{t_0} &=
      \<if>(\compileExp{b}{t_0}, \compileExp{e_0}{t_0}, \compileExp{e_1}{t_0})\\
      \compileExp{\<obs>(l,i)}{t_0} &= \<model>(l, \<tplus>(t_0,i))\\
      \compileContr{\<transfer>(p_1,p_2,a)}{t_0} &=  \<payoff>(t_0,p_1,p_2)\\
      \compileContr{\<scale>(e,c)}{t_0} &= \<mult>(\compileExp{e}{t_0}, \compileContr{c}{t_0})
 \end{aligned}
\]
\[
\begin{aligned}[t]
  \compileContr{\<zero>}{t_0} &= 0\\
  \compileContr{\<translate>(t,c)}{t_0} &= \compileContr{c}{\smartTplus(t_0, t)}\\
  \compileContr{\<both>(c_0,c_1)}{t_0} &= \<add>(\compileContr{c_0}{t_0}, \compileContr{c_1}{t_0})\\
\end{aligned}
\]
\begin{multline*}
  \compileContr{\<ifWithin>(e,t,c_1,c_2)}{t_0} =\\
  \<loopif>(\compileExp{e}{t_0},\compileContr{c_0}{t_0},\compileContr{c_1}{t_0}, t)
\end{multline*}
\[
\begin{aligned}[t]
  \smartTplus(t_1, t_2) &=
  \begin{cases}
    t_1 + t_2 ~ \text{if } t_1,t_2 ~ \text{ are numerals}\\
    \<tplus>(t_1,t_2) ~ \text{otherwise}
  \end{cases}
\end{aligned}
\]
The important point to notice here is that all relative time shifts in
an expression in CL are accumulated to the $t_0$ parameter. The
resulting payoff expression only contains lookups in the external
environment where time is given explicitly, and does not depend on
nesting of time shifts as it was in the case of $\transl{t}{c}$ in
CL. Such a representation allows for a more straightforward evaluation
model.\footnote{The $\mathtt{loopif}$ construct still introduces
  relative time shifts similarly to $\mathtt{ifWithin}$. This makes
  code generation in a target language less trivial. Potentially, it
  is possible to decompose $\mathtt{loopif}$ into an iteration and a
  conditional expression. Also, in the case when number of iterations
  of $\mathtt{loopif}$ is a number and not a template variable, it is
  possible to completely unroll $\mathtt{loopif}$ into nested
  conditional expressions, making all the indexing into the external
  environment explicit.} We also would like to emphasise that $\acc$
and $\letA$ constructs are not supported by our compilation
procedure. On the supported subset of the contract language,
the compilation functions $\compileExp{e}{}$, and $\compileContr{c}{}$ are
total. We use the \emph{smart constructor} $\smartTplus$ that adds two
integer literals whenever it is possible or returns a syntactic
addition expression. This use of smart constructors is useful not only for optimisation
purposes; it also allows us to overcome some difficulties in
formalisation (see Remark~\ref{rem:contracts:templ-expr-compile} in
Section~\ref{sec:contracts:coq-formalisation}).

\begin{example}\label{ex:contracts:compile}
We consider the following contract (\icode{t0} and \icode{t1} denote template
variables): the party ``you'' transfer to the party ``me'' 100 USD in \icode{t0}
days in the future, and after \icode{t1} more days ``you'' transfers to ``me''
an amount equal to the difference between the current price of the AAPL stock and 100
USD, provided that the price of AAPL is higher then 100 USD
(we use infix notation for arithmetic operations to make code more readable).
\begin{lstlisting}
 c $=$
   translate(t0,
   both(scale(100.0, transfer(you,me)),
        translate(t1,
          if(obs(AAPL,0) > 100.0,
             scale(obs(AAPL,0) - 100.0, transfer(you, me)),
             zero)))
\end{lstlisting}

This contract compiles to the following code in the payoff intermediate language:
\begin{lstlisting}
 e $=$
   (100.0 * payoff(t0,you,me)) +
   if (model(AAPL,t0+t1) > 100.0,
       (model(AAPL,t0+t1) - 100.0) * payoff(t0+t1,you,me),
       0.0)
\end{lstlisting}
As one can see, all nested occurrences of the \icode{translate} construct were
accumulated from top to bottom. That is, in the \icode{if} case, we calculate payoffs
and lookup for values of the AAPL stock at time \icode{(t0+t1)}.
\end{example}

To be able to reason about soundness of the compilation process, one
needs to make a connection between the semantics of the two
languages. For the expression sublanguage of CL ($e \in \type{Exp}$)
we can just compare the values that the original
expression and the compiled expression evaluates to. In case of the
contract language ($c \in \type{Contr}$) the situation is different,
since the semantics of a contract is given in terms of a
$\type{Trace}$, and an expression in the payoff intermediate language
evaluates to a single value. However, we know that the compiled
expression represents the sum of the contract cashflows after
discounting.

We assume a function $\id{HOR}: \type{TEnv}\times\type{Contr} -> \mathbb{N}$
that returns a conservative upper bound on the length of a contract. We often
write $\compileExp{e}0 = \id{il}$, or $\compileContr{c}0 = \id{il}$ to emphasise
that the compilation function returns some result. The value environment
$\gSem{\Gamma}$ is not relevant for the present development and we will omit
it. The compilation function satisfies the following properties:

\begin{theorem}[Soundness]\label{thm:contracts:compile-sound}
  Assume parties $p_1$ and $p_2$ and discount function $d : \mathbb{N} -> \mathbb{R}$, environments
  $\rho \in \type{Env}$, and $\delta \in \type{TEnv}$.
  \begin{enumerate}[(i)]
    \item If ~$\compileExp{e}0 = \id{il}$ and
      $\eSem{e}{\rho,\delta} = v_1$ and $\ilSem{\id{il}}{\rho,\delta,0,0,d,p_1,p_2}
      = v_2$ then $v_1=v_2$.

    \item  If ~$\compileContr{c}0 = \id{il}$ and ~$\cSem{c}{\rho,\delta} = \id{tr}$,
      where $\id{tr} : \mathbb{N} -> \type{Party} \times \type{Party} -> \mathbb{R}$
      then
      \begin{center}
        $\sum_{t=0}^{\synhor_\delta(c)} d(t) \times \id{tr}(t)(p_1,p_2) = \ilSem{\id{il}}{\rho,\delta,0,0,d,p_1,p_2}$
      \end{center}
  \end{enumerate}
\end{theorem}

Theorem \ref{thm:contracts:compile-sound} makes an assumption that the
compiled expression evaluates to some value. We do not develop a type
system for our payoff language to ensure this property. Instead, we
show that it is sufficient for a contract to be well-typed to ensure
that the compiled expression always evaluates to some value (for
details, we refer the reader to the typing rules for the contract
language in \cite{BahrBertholdElsman}).

\begin{theorem}[Total semantics for compiled contracts]\label{thm:contracts:well-typed-compiled}
  Assume parties $p_1$ and $p_2$ and discount function $d : \mathbb{N}
  -> \mathbb{R}$, well-typed external environment $\rho \in
  \type{Env}$, template environment $\delta \in \type{TEnv}$, and
  typing context $\Gamma$.  The following two properties hold:
  \begin{enumerate}[(i)]
  \item for any
    $e \in \type{Exp}$, $t_0 \in \type{TExprZ}$, $t_0' \in \nats$, if
    $\Gamma \vdash e : \tau$
    $\compileExp{e}{t_0} = il$, then
    \[\exists v,~\ilSem{\id{il}}{\rho,\delta,t_0',0,d,p_1,p_2} = v
    \text{, ~and~} v \in \tysem{\tau}\]
  \item for any
    $c \in \type{Contr}$, $t_0$ $t_0'$, if $\Gamma \vdash c$ and $\compileContr{c}{t_0} = il$, then
    \[\exists v,~\ilSem{\id{il}}{\rho,\delta,t_0',0,d,p_1,p_2} = v  \text{, ~and~} v \in \reals\]
  \end{enumerate}
\end{theorem}

Notice that Theorem~\ref{thm:contracts:well-typed-compiled} holds for
any $t_0 \in \type{TExprZ}$ and $t_0' \in \nats$.  These parameters do
not affect totality of the semantics and can be arbitrary, since we
assume that the external environment is
total. Theorems~\ref{thm:contracts:compile-sound}
and~\ref{thm:contracts:well-typed-compiled} together ensure that our
compilation procedure produces a payoff expression that evaluates to a
value reflecting the aggregated price of a contract after discounting.

\subsection{Avoiding recompilation}
To avoid recompilation of a contract when time moves forward, we define a
function $\cutPayoff{}$. This function is defined recursively on the syntax of
intermediate language expressions.
\begin{multline*}
\begin{aligned}
  \mathsf{cutPayoff} &: \type{ILExpr} -> \type{ILExpr}\\
  \cutPayoff{\<now>} &= \<now>\\
  \cutPayoff{\<model>(l,t)} &= \<model>(l,t)\\
  \cutPayoff{\id{unop}(\id{il})} &= \id{unop}(\cutPayoff{\id{il}}) \\
\end{aligned}\\[0.5em]
\cutPayoff{\<payoff>(t,p_1,p_2)} = \\
\<if>(t < \<now>,0,\<payoff>(t,p_1,p_2))\\[0.5em]
\cutPayoff{\id{binop}(\id{il}_1,\id{il}_2)} =\\
\id{binop}(\cutPayoff{\id{il}_1},\cutPayoff{\id{il}_2})\\[0.5em]
\cutPayoff{\<if>(\id{il}_1,\id{il}_2,\id{il}_3)} =\\
\<if>(\cutPayoff{\id{il}_1},\cutPayoff{\id{il}_2},\cutPayoff{\id{il}_3})
\end{multline*}
The most important case is the case for $\<payoff>$. The function wraps
$\<payoff>$ with a condition guarding whether this payoff affects the resulting
value. For the remaining cases, the function recurses on subexpressions and
returns otherwise unmodified expressions.

\begin{example}
  Let us consider Example \ref{ex:contracts:compile} again and apply the $\cutPayoff$
  function to the expression \icode{e}:
  \begin{lstlisting}
 cutPayoff(e) $=$
   (100.0 * disc(t0) * if(t0 < now, 0, payoff(you,me)) +
   if (model(AAPL,t0+t1) > 100.0,
       (model(AAPL,t0+t1) - 100.0) * disc(t0+t1) *
        if(t1+t0 < now, 0, payoff(you,me)),
       0.0)
  \end{lstlisting}

  Each \icode{payoff} in the payoff expression is now guarded by the condition,
  comparing the time of the particular payoff with \icode{now}. Notice that the
  templates variables \icode{t0} and \icode{t1} are mapped to concrete values in
  the template environment.
\end{example}

To be able to state a soundness property for the $\cutPayoff$ function we again
need to find a way to connect it to the semantics of CL. Since
$\cutPayoff$ deals with the dynamic behavior of the contract with respect to
time, it seems natural to formulate the soundness property in this case in
terms of contract reduction (\cite[Figure~10]{BahrBertholdElsman}). The semantics of the
payoff language takes the ``current time'' $t$ as a parameter. We should be
able to connect the $t$ parameter to the step of contract reduction.

\begin{theorem}[Contract compilation soudness wrt. contract reduction]\label{thm:contracts:soundness-red}
  We assume parties $p_1$, $p_2$, discount function $d : \nats -> \reals$.
  For any well-typed and template-closed contract $c$, i.e., we assume $\Gamma |-
  c$, and $\TC{c}$, an external environment $\rho' \in \type{Env}$ extending a
  partial external environment $\rho \in \type{Env_p}$, if $c$ steps to some
  $c'$ by the reduction relation $c ~\cRed{\rho}{T}~c'$, for some transfer
  $T \in \type{Trans}$, such that $\cSem{c'}{(\rho'/1),\emptyset} = \id{trace}$, and $\compileContr{c}0=il$,
  then
  \[
  \sum_{t'=0}^{\synhor_\delta(c')} d(t'+1) \times \id{trace}(t') =
  \ilSem{\cutPayoff{il}}{\rho',\emptyset,0,1,d,p_1,p_2}
  \]
  where $\rho'/1$ denotes the external environment $\rho$ advanced by one time
  step:
  \[ \rho'/1 = \lambda (l,i).~\rho'(l,i + 1), \quad l \in \type{Label}, i \in \ints \]
\end{theorem}

From the contract pricing perspective, the partial external environment $\rho$
contains \emph{historical data} (e.g., historical stock quotes) and the extended
environment $\rho'$ is a union of the two environments $\rho$ and $\rho''$, where
$\rho''$ contains \emph{simulated data}, produced by means of simulation in
the pricing engine (e.g., using Monte Carlo techniques).

Avoiding recompilation can significantly improve performance especially on GPGPU
devices. On the other hand, additional conditionals are introduced, which
results in a number of additional checks at run-time. We have investigated the
influence on performance of these additional conditions for certain
contracts. The results of experiments are given in Section~\ref{subsec:conracts:futhark}.

One also might be interested in the following property. The following two ways
of using our compilation procedure give identical results:
\begin{itemize}
\item first reduce a contract (move time forward), compile it to a payoff expression, then evaluate the payoff expression;
\item first compile a contract to a payoff expression, apply $\cutPayoff$ to the payoff expression, and then evaluate, specifying the appropriate
  value for the ``current time'' parameter.
\end{itemize}

Let us introduce some notation first.  We fix the well-typed external
environment $\rho$, the partial environment $\rho'$, which is historically
complete ($\rho'(l,i)$ is defined for all labels $l$ and $i \leq 0$), and a
discount function $d : \nats -> \reals$. Next, we assume that contracts are
well-typed, and closed with respect to template variables, the compilation function
is applied to supported constructs only, and that the reduction function,
corresponding to the reduction relation, is total on $\rho'$ (see \cite[Theorem
  11]{BahrBertholdElsman}). This gives us the following total functions:
\[
\begin{aligned}
    red_{\rho'} : \type{Contr} -> \type{Contr}\\
    \compileContr{-}0 : \type{Contr} -> \type{ILExpr}\\
  \end{aligned}
  \]
  These functions correspond to the contract reduction function and the contract compilation function.
  We also define an evaluation function for compiled payoff expressions as a shortcut
  for the payoff expression semantics.
\[
\begin{aligned}
    \id{evalAt}_{-} &: \nats -> \type{ILExpr} \times \type{Env} \times \type{Disc} -> \reals + \bools\\
    \id{evalAt}_t (e,\rho,d) &= \ilSem{e}{\rho,\emptyset,0,t,d,p_1,p_2}
\end{aligned}
\]
for some parties $p_1$ and $p_2$. We know by Theorem
\ref{thm:contracts:well-typed-compiled} that $\id{evalAt}$ is total on payoff
expressions produced by the compilation function from well-typed contracts.

We summarise the property by depicting it as a commuting diagram.
\begin{theorem}\label{thm:contracts:commutes}
  The following diagram commutes:
\begin{center}
\begin{tikzcd}
  \type{Contr} \arrow[rr, "\id{red}_{\rho'}"] \arrow[d, "\mathsf{cutPayoff}~\circ~\compileContr{-}0",swap]
    &  & \type{Contr} \arrow[d, "\compileContr{-}0"] \\
    \type{ILExpr} \arrow[rd,"\id{evalAt}_1\tuple{-,\rho,d}",swap]
    & & \type{ILExpr} \arrow[ld,"\id{evalAt}_0\tuple{-,\rho/1,d/1}"] \\
    & \reals &
\end{tikzcd}
\end{center}
Here we write $\rho/1$ and $d/1$ for shifted one step external environment and
discount function, respectively.
\end{theorem}
The above diagram gives rise to the following equation:
\begin{multline*}
  \id{evalAt}_1\tuple{-,\rho,d} \circ \mathsf{cutPayoff}~\circ~\compileContr{-}0 =\\
\id{evalAt}_0\tuple{-,\rho/1,d/1} \circ \compileContr{-}0 \circ \id{red}_{\rho'}
\end{multline*}
 This property shows that our implementation can be used in two
 different ways. Either a contract is compiled upfront with
 $\cutPayoff$ or a contract is reduced, at each time of interest, to
 another contract, which is then compiled to a payoff expression for
 evaluation. The second use case allows for more flexibility for
 users. For example, one can develop a system where users define
 contracts directly in terms of CL working in a specialised IDE. The
 first case allows for compiling upfront a set of predefined
 financial instruments (or contract templates) avoiding recompilation
 when time moves forward. Adding a new instrument is possible, but
 requires recompilation.

The statements of Theorems~\ref{thm:contracts:soundness-red}
and~\ref{thm:contracts:commutes} generalise in the obvious way to $n$-step
reduction (by replacing one-step reduction with $n$-step reduction,
replacing environment shifts from $\rho/1$ to $\rho/n$, and
evaluating the compiled payoff expression at $t=n$ instead of $t=1$).  The
crucial step for proving these generalised theorems is to use the
following theorem.

\begin{theorem}[Soundness of $\cutPayoff$ for $n$ time steps]\label{thm:contracts:nstep}
  Assume parties $p_1$ and $p_2$, a discount function $d : \mathbb{N} -> \mathbb{R}$, and
  a well-typed external environment $\rho \in \type{Env}$.
  For any well-typed and template-closed contact $c$ at a time step $n \in \nats$,
  if ~$\compileContr{c}0 = \id{il}$ and ~$\cSem{c}{\rho,\emptyset} = \id{tr}$ then
  \begin{center}
    $\sum_{t=n}^{\synhor_\delta(c)} d(t) \times \id{tr}(t)(p_1,p_2) = \ilSem{\cutPayoff{il}}{\rho,\emptyset,0,n,d,p_1,p_2}$
  \end{center}
\end{theorem}
Theorem~\ref{thm:contracts:nstep} expresses soundness of $n$-step
$\cutPayoff$ evaluation without explicitly mentioning contract
reduction and leads to a better proof structure in the Coq
formalisation. Intuitively, this theorem says that the sum of the trace
starting at $n$ instead of zero is exactly the value we obtain after
evaluating $\cutPayoff{il}$ at current time $t=n$.

Theorem~\ref{thm:contracts:nstep} combined with the properties of $n$-step reduction gives us the proofs of Theorems~\ref{thm:contracts:soundness-red} and~\ref{thm:contracts:commutes}. Our Coq development
(see Section \ref{sec:contracts:coq-formalisation}) contains a full formalisation of these generalised theorems.

\section{Formalisation in Coq}\label{sec:contracts:coq-formalisation}
\lstset{language=Coq}

Our formalisation in Coq\footnote{The formalisation presented in this
  paper is available online:
  \url{https://github.com/annenkov/contracts}. The repository includes
  the backends generating Haskell and Futhark code along with the
  pricing engine implementation in Futhark for benchmarking.} extends
the previous work \cite{BahrBertholdElsman} by introducing the concept
of template expressions and by developing a certified compilation
technique for translating contracts to payoff expressions. The
required modifications to the denotational semantics have been
presented in Section~\ref{subsec:contracts:contracts-syn-sem}. These
modifications required us to propagate changes to all the proofs
affected by the change of syntax and semantics. We start this section
with a description of the original formalisation, and then continue
with modifications and additions made by the authors of this work.

The formalisation described in \cite{BahrBertholdElsman} uses an extrinsic
encoding of CL. That means that syntax is represented using Coq's
inductive data types, and a typing relation on these \emph{raw} terms are
given separately. For example, the type of the expression sublanguage is
defined as follows.
\begin{lstlisting}[language=Coq]
   Inductive Exp : Set :=
     OpE (op : Op) (args : list Exp)
   | Obs (l : ObsLabel) (i : Z)
   | VarE (v : Var)
   | Acc (f : Exp) (d : nat) (e : Exp).
\end{lstlisting}

One of the design choices in the definition of \icode{Exp} is to make the
constructor of operations \icode{OpE} take ``code'' for an operation
and the list of arguments. Such an implementation makes adding new operations
somewhat easier. Although, we would like to point out that this definition
is a \emph{nested} inductive definition (see \cite[Section~3.8]{cpdt}). In such
cases Coq cannot automatically derive a strong enough induction principle,
which means that it needs to be defined manually. In the case of \icode{Exp} it is not hard to
see, that one needs to add a generalised induction hypothesis in case of
\icode{OpE}, saying that some predicate holds for all elements in the argument
list.

Although the extrinsic encoding requires more work in terms of proving, it has a
big advantage for code extraction, since simple inductive data types are easier
to use in the Haskell wrapper for CL.

One of the consequences of this encoding is that semantic functions for
contracts $\type{Contr}$ and expressions $\type{Exp}$ are partial, since they
are defined on raw terms which may not be well-typed.  This partiality is
implemented with the \icode{Option} type, which is equivalent to Haskell's
\icode{Maybe}. To structure the usage of these partial functions, we
define the \icode{Option} monad and use monadic binding
\begin{lstlisting}
   bind : forall A B : Type,
          option A -> (A -> option B) -> option B
\end{lstlisting}
to compose calls of partial functions together. The functions
\begin{lstlisting}
   liftM: forall A B : Type,
          (A -> B) -> option A -> option B
   liftM2 : forall A B C : Type,
            (A -> B -> C) -> option A -> option B
            -> option C
   liftM3 : forall A B C D : Type,
            (A -> B -> C -> D) -> option A -> option B
            -> option C -> option D
\end{lstlisting}
allow for a total function of one, two, or three arguments to be lifted to the
\icode{Option} type. The implementation includes poofs of some properties of
\icode{bind} and the lifting functions. These properties include cases for which an
expression evaluates to some value.
\begin{lstlisting}
bind_some : forall (A B : Type) (x : option A)
                  (v : B) (f : A -> option B),
              x >>= f = Some v
              -> exists x' : A, x = Some x' /\ f x' = Some v
\end{lstlisting}

Similar lemmas were proved for other functions related to the
\icode{Option} type. To simplify the work with the \icode{Option} monad, the
implementation defines tactics in the \texttt{Ltac} language (part of Coq's
infrastructure). The tactics \icode{option_inv} and \icode{option_inv_auto} use
properties of operations like \icode{bind} and \icode{liftM} to invert
hypotheses like \icode{e = Some v}, where $e$ contains the aforementioned
functions. The implementation uses some tactics from
\cite{pierce}. Particularly, the \icode{tryfalse} tactic is widely used.  It
tries to resolve the current goal by looking for contradictions in assumptions,
which conveniently removes impossible cases.

The original formalisation of the contract language has been modified by introducing the type of
template expressions
\begin{lstlisting}
   Parameter TVar : Set.
   Inductive TExpr : Set :=
       Tvar (t : TVar)
     | Tnum (n : nat).
\end{lstlisting}

We keep the type of variables abstract and do not impose any restrictions on
it. Although one could add decidability of equality for \icode{TVar}, if
required, we do not compare template variables in our formalisation.
We modify the definition of the type of contracts \icode{Contr} such that
constructors of expressions related to temporal aspects now accept \icode{TExpr}
instead of \icode{nat} (\icode{If} corresponds to $\<ifWithin>$):
\begin{lstlisting}
   Translate : TExpr -> Contr -> Contr
   If : Exp -> TExpr -> Contr -> Contr -> Contr.
\end{lstlisting}
We leave the other constructors unmodified.

\begin{sloppypar}
Similarly to how we define an external environment, we define a
\emph{template environment} as a function type \icode{TEnv := TVar ->
  nat}. Such a definition allows for easier modification of existing
code base in comparison with partial mappings. According to the
definitions in Section~\ref{subsec:contracts:contracts-syn-sem}, we
modify the semantic function for contracts, and the symbolic horizon
function, to take an additional parameter of type
\icode{TEnv}. Propagation of these changes was not very problematic
and almost mechanical. Because the first attempt to parameterise the
reduction relation with a template environment led to some problems,
we decided to define the reduction relation only for
template-closed contracts.  In most cases it is sufficient to
instantiate a contract, containing template variables using the
instantiation function (Definition~\ref{eq:contratcs:inst-func}), and
then reduce it to a new contract. Although instantiation requires a
template environment containing all the mappings for template variables
mentioned in the contract, we do not consider this a big limitation.
\end{sloppypar}

The definition of the payoff intermediate language (following
Section~\ref{subsec:contracts:payoffs-syn-sem}) also uses an extrinsic
encoding to represent raw terms as an inductive data type. We define
one type for the payoff language expressions \icode{ILExpr}, since
there is no such separation as in CL on contracts and expressions. The
definition of template expressions used in the definition of
\icode{ILExpr} is an extended version of the definition of template
expressions \icode{TExpr} used in the contract language definition.
\begin{lstlisting}
   Inductive ILTExpr : Set :=
     ILTplus (e1 : ILTExpr) (e2 : ILTExpr)
   | ILTexpr (e : TExpr).

   Inductive ILTExprZ : Set :=
     ILTplusZ (e1 : ILTExprZ) (e2 : ILTExprZ)
   | ILTexprZ (e : ILTExpr)
   | ILTnumZ (z : Z).
\end{lstlisting}

Notice that we use two different types of template expressions \icode{ILTExpr}
and \icode{ILTExprZ}. The former extends the definition of \icode{TExpr} with
the addition operation, and the latter extends it further with integer literals
and with the corresponding addition operation (recall that template expressions
used in CL can be either natural number literals or variables).  The
reason why we have to extend \icode{TExpr} with addition is that we want to
accumulate time shifts introduced by \icode{Translate} in one expression using
(syntactic) addition. In the expression sublanguage of CL, observables
can refer to the past by negative time indices. For that reason we introduce
the \icode{ILTExprZ} type.

The full definition of syntax for the payoff intermediate language in our Coq
formalisation looks as follows:
\begin{lstlisting}
   Inductive ILExpr : Set :=
   | ILIf : ILExpr -> ILExpr -> ILExpr -> ILExpr
   | ILFloat : R -> ILExpr
   | ILNat : nat -> ILExpr
   | ILBool : bool -> ILExpr
   | ILtexpr : ILTExpr -> ILExpr
   | ILNow  :  ILExpr
   | ILModel : ObsLabel -> ILTExprZ -> ILExpr
   | ILUnExpr : ILUnOp -> ILExpr -> ILExpr
   | ILBinExpr : ILBinOp -> ILExpr -> ILExpr -> ILExpr
   | ILLoopIf : ILExpr -> ILExpr -> ILExpr -> TExpr -> ILExpr
   | ILPayoff  : ILTExpr -> Party -> Party -> ILExpr.
\end{lstlisting}
Notice that we use template expressions, which could represent negative
numbers (\icode{ILTExprZ}) in the constructor \icode{ILModel}. This
constructor corresponds to observable values in the contract language
and allows for negative time indices corresponding to access of historical data.

We could have generalised our formalisation to deal with different types of
template variables and added a simple type system on top of the template
expression language, but we decided to keep our implementation simple, since
the main goal was to demonstrate that it is possible to extend the original
contract language to contract templates with temporal variables.

All the theorems and lemmas described in the paper are
completely formalised in our Coq development. We use a limited amount of proof
automation in the soundness proofs. The proof automation is used
mainly in the proofs related to compilation of the
contract expression sublanguage, since compilation is straightforward and
proofs are relatively easily to automate. Moreover, without the proof automation,
one would have to consider a large number of very similar cases leading to code
duplication. In addition to \icode{option_inv_auto} mentioned above, we use a
tactic that helps to get rid of cases where expressions (a source expression in
\icode{Exp} and a target expression in \icode{ILEpxr}) evaluate to values of
different types (denoted by the corresponding constructor).
\begin{lstlisting}
   Ltac destruct_vals :=
     repeat (match goal with
             | [x : Val |- _] => destruct x; tryfalse
             | [x : ILVal |- _] => destruct x; tryfalse
             end).
\end{lstlisting}
Here the \icode{Val} and \icode{IVal} types correspond to values of the
contract expression sublanguage and the payoff expression language
respectively. The \icode{tryfalse} tactic searches for the contradictions in the
goal (see \cite{pierce}).

Another tactic that significantly reduces the complexity of the proofs is the
\icode{omega} tactic from Coq's standard library. This tactic implements a
decision procedure for expressions in Presburger arithmetic. That is, goals can
be equations or inequations of integers, or natural numbers with addition and
multiplication by a constant. The tactic uses assumptions from the current
context to solve the goal automatically.

The principle we use in the organisation of the proofs is to use proof automation
to solve the most trivial and tedious goals and to be more explicit about the proof
structure in cases requiring more sophisticated reasoning.

\begin{remark}\label{rem:contracts:templ-expr-compile}
  The first version of the soundness proof was developed for the original
  contract language without template expressions. The proof was somewhat easier,
  since the aggregation of nested time shifts introduced by $\transl n c$
  constructs during compilation was implemented as addition of natural numbers,
  corresponding to time shifts. In the presence of template expressions, the
  compilation function builds a syntactic expression using the $\<tplus>$
  constructor. There are some places in proofs where it was crucial to use
  associativity of addition to prove the goal, but this does not work for
  template expressions. For example, $\<tplus>(\<tplus>(t_1,t_2),t_3)$ is not
  equal to $\<tplus>(t_1,\<tplus>(t_2,t_3))$, because these expressions
  represent different syntactic trees, although semantically
  equivalent. Instead of restating proofs in terms of this semantic equivalence
  (significantly complicating the proofs), we used the following approach.  The
  compilation function uses the smart constructor $\smartTplus$ instead
  of just plain construction of the template expression. This allowed us to
  recover the property we needed to complete the soundness proof without
  altering too much of its structure.
\end{remark}

There are a number of aspects that introduce complications to the development of proofs
of the compilation properties.
\begin{itemize}
  \item Accumulation of relative time shifts during compilation. To
    obtain a general enough induction hypothesis we have to generalise
    our lemmas to take as parameter an initial time \icode{t0}. The
    same holds for the semantics of \icode{loopif}, since there is an
    additional parameter in the semantics to implement iterative
    behavior.
  \item Presence of template expressions. The complications we faced
    due to template expressions are described in
    Remark~\ref{rem:contracts:templ-expr-compile}. We have resolved
    these complications with smart constructors, but template
    expressions still add some overhead.
  \item Conversion between types of numbers. We use integers and natural numbers
    (\icode{nat} and \icode{Z} type from the standard library of Coq). In some
    places, including the semantics of template expressions, we use a conversion
    from natural numbers to integers. This conversion makes automation with the
    \icode{omega} tactic more complicated, because it requires first to use the
    properties of conversion, which is harder to automate. With the accumulation
    aspect, conversions add even more overhead.
  \item We use the definition of contract horizon in the statement of
    the soundness theorems, which leads to additional case analysis in
    proofs.
\end{itemize}

\subsection{Code Extraction}\label{subsec:contracts-code-extraction}
The Coq proof assistant allows for extracting Coq functions into programs in
some functional languages \cite{coqextract2008}. The implementation described
in \cite{BahrBertholdElsman} supports code extraction of the contract
type checker and contract manipulation functions into the Haskell programming
language. We extend the code extraction part of the implementation with
features related to contract templates and contract compilation. Particularly,
we extract Haskell implementations of the following functions:
\begin{itemize}
\item \icode{inst_contr} function that instantiates a given contract according to given
  template environment;
\item \icode{fromExp} function for compiling the contract expression sublanguage;
\item \icode{fromContr} function for compiling contract language constructs;
\item \icode{cutPayoff} function for parameterising a payoff expression with
  the ``current time''.
\item \icode{ILsem} semantic function for payoff expressions, which can be
  used as an interpreter.
\end{itemize}

For supporting templates, we have updated the Haskell front end and
exposed the full contract language in a convenient form. We have kept
the original versions of extended combinators, such as
\icode{translate} and \icode{within} without changes and added
\icode{translateT} and \icode{withinT} combinators, which support
template variables.

Our implementation contains an extended collection of contract examples, examples
of contract compilation, and evaluation of resulting payoff expressions.

\section{Code Generation}\label{subsec:contracts:codegen}
To exemplify how the payoff language can be used to produce a payoff function
in a subset of some general purpose language, we have implemented a code
generation procedure to Haskell and Futhark, as illustrated in the following diagram:
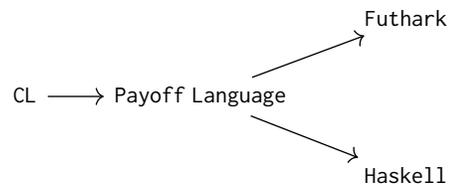
\begin{center}
  \begin{tikzcd}
    & & \type{Futhark}\\
  \type{CL} \arrow[r, ""] & \type{Payoff~Language} \arrow[ru,""] \arrow[rd,""] \\
    & & \type{Haskell}
\end{tikzcd}
\end{center}
We make use of the code extraction mechanism described in
Section~\ref{subsec:contracts-code-extraction} to obtain a certified
compilation function, which we use to translate expressions in CL to
expressions in the payoff language.

\subsection{The Haskell Backend}\label{subsec:contracts:haskell}
The code generation procedure is (almost) a one-to-one mapping of the
payoff language constructs to Haskell expressions. One primitive,
which we could not map directly to Haskell build-in functions was the
\icode{loopif} construct. We have solved this issue by implementing
\icode{loopif} as a higher-order function in Haskell. The
implementation essentially follows the definition of the semantics of
\icode{loopif} in Coq:

\begin{lstlisting}[language=Haskell]
loopif :: Int -> Int -> (Int -> Bool) -> (Int -> a) -> (Int -> a) -> a
loopif n t0 b e1 e2 = let b' = b t0 in
                      $~$case b' of
                        True -> e1 t0
                        False -> case n of
                                  0 -> e2 t0
                                  _ -> loopif (n$-$1) (t0+1) b e1 e2
\end{lstlisting}

\noindent
The resulting payoff function has the following signature:
\begin{lstlisting}[language=Haskell]
   payoff :: Map.Map ([Char], Int) Double -> Map.Map [Char] Int
             -> Int -> Party -> Party -> Double
\end{lstlisting}
That is, the function takes as parameters an external environment, a template environment,
the current time, and two parties. The \icode{payoff} function calls the
\icode{payoffInternal} function, which takes an additional parameter---an
initial value for the \icode{loopif} function, which serves as a loop counter.

\begin{example}
  We apply the code generation procedure to the expression \icode{e}
  from Example~\ref{ex:contracts:compile}. Here is the result of code
  generation:
  \begin{lstlisting}[language=Haskell]
module Examples.PayoffFunction where
import qualified Data.Map as Map
import BaseTypes
import Examples.BasePayoff

payoffInternal ext tenv t0 t_now p1 p2 =
   (100.0 * (if  (X== p1 && Y== p2) then 1
             else if (X== p2 && Y== p1) then $-$1 else 0)) +
   (if ((100.0 < (ext Map.! ("AAPL",(0 + (tenv Map.! "t1") +
                                  (tenv Map.! "t0") + 0+ t0)))))
    then ((((ext Map.! ("AAPL",(0 + (tenv Map.! "t1") +
                              (tenv Map.! "t0") + 0+ t0))) * 100.0)
           * (if  (X== p1 && Y== p2) then 1 else
              if  (X== p2 && Y== p1) then $-$1 else 0))) else 0.0)

payoff ext tenv t_now p1 p2 = payoffInternal ext tenv 0 t_now p1 p2
  \end{lstlisting}

  The external environment and the template environment are
  represented using Haskell's \icode{Data.Map}, and \icode{Map.!} is an infix
  notation for the lookup function. To obtain the code above we apply a simple
  optimisation, replacing the \icode{loopif} with zero as the first argument
  with the regular \icode{if}. One could also add more optimisations to our Coq
  implementation along with proofs of soundness.

  A module declaring the \icode{payoff} function can be used as an ordinary
  Haskell module as a part of the development requiring the payoff
  functions. For example, it could be used in the context of the FinPar
  benchmark \cite{Andreetta:2016:FPF:2952301.2898354}, which contains a Haskell
  implementation of pricing among other routines.  Moreover, the $\cutPayoff{}$
  function can be used to obtain a parameterised version of a payoff function
  in Haskell, allowing us to reproduce the contract reduction behavior.
\end{example}

\subsection{The Futhark Backend}\label{subsec:conracts:futhark}
Futhark is a data-parallel functional language for programming nested,
regular programs to be executed efficiently on a GPU
\cite{henriksen2014size,Henriksen:Futhark}. The language has a rich
core language, which provides a number of second-order functional
array combinators, such as \icode{map}, \icode{reduce},
\icode{filter}, and \icode{scan}, but it also provides seemingly
imperative features, including sequential loops and array updates,
which are based on a uniqueness type system that allows for an
efficient implementation of functional array updates. On top of the
core language, Futhark is enriched with a higher-order module
language, for which constructs are compiled away at compile time due
to a static interpretation technique.

Generating code for Futhark is quite similar to the code generation
approach described for Haskell. With the Futhark backend, however, the
aim is to integrate generated payoff functions with an efficient
parallel Monte Carlo based pricing engine, which is achieved by making
the pricing engine a parameterised module that takes as argument a
module containing a payoff-function.

Regarding the particular Futhark payoff function generation, the
implementation differs from the Haskell implementation in two ways,
namely (i) with respect to the representation of the $\mathtt{loopif}$
construct and (ii) with respect to external environment access.

The first difference is related to the fact that Futhark does not
support recursive functions, but instead includes various iteration
constructs. The payoff language $\mathtt{loopif}$ construct is
therefore compiled into a Futhark $\mathtt{while}$ loop construct. For
example, consider the following payoff expression (corresponding to a
simple contract with a barrier):
\begin{lstlisting}[language=Contracts]
   loopif(model("AAPL", 0) <= 4000.0,
           0.0, 2000.0 * payoff(0,X,Y), t)
\end{lstlisting}
This payoff expression is translated into the following fragment of
Futhark code:
\begin{lstlisting}[language=Futhark]
   let payoffInternal(ext : [][]f32, tenv : []i32,
                      disc : []f32, t0 : i32, t_now : i32) : f32 =
     let t0 = loop t0 = t0
         while (!(ext[t0,0] <= 4000.0) && (t0 < tenv[0])) do t0+1
         in if (ext[t0,0] <= 4000.0)
            then 0.0
            else (2000.0 * disc[t0])

   let payoff ext tenv t_now = payoffInternal(ext,tenv,0,t_now)
\end{lstlisting}

\noindent
The \icode{ext} variable is a two-dimensional array containing model data (the
first index corresponds to time and the second corresponds to an observable),
\icode{tenv} is an array with template parameter values, and \icode{disc} is an
array containing discount factors (indexed by time).

The second difference, which is related to the way we work with the
model data environment, is concerned with translation of environment
indexing to the form used in the FinPar pricing code. For example, we
translate time indices 100 and 200 in the following payoff expression
\begin{lstlisting}[language=Contracts]
   payoff(100,X,Y) + payoff(200,X,Y)
\end{lstlisting}
\noindent
to 0 and 1 respectively. This reindexing corresponds to the order in which time
indices appear in a payoff expression.

The output of the generation procedure is a Futhark module that can be
directly passed to the parameterised Futhark pricing engine module. A
key feature of the implemented template mechanism combined with the
$\cutPayoff{}$ functionality is that the code base needs to be compiled
into efficient GPU code only when new instruments are introduced; the
generated payoff expressions are generic with respect to the time at
which the price is calculated.



Table~\ref{timings.tab} shows the timings for pricing three different
financial contracts using the FinPar Monte Carlo pricing engine
\cite{Andreetta:2016:FPF:2952301.2898354}. The contracts include a
\emph{vanilla European call option}, which allows a holder at some
time $t$ to purchase a particular stock at a predetermined price, and
a \emph{discrete barrier option}, which forces a holder to exercise
the option before maturity if any of three particular underlying
stocks at certain dates cross certain barrier levels. Finally, the
contracts include a \emph{double vanilla European option}, which
allows a holder to exercise any of two European options on two
different underlying stocks. The three contracts cover well the
possible scope of supported contracts, including the support for
dealing with multiple underlyings and multiple measurement
days. Moreover, the contracts are instances of real financial
contracts appearing in real financial portfolios.

The experiments were executed on a
commodity MacBook Pro laptop with a 2.7GHz Intel i7 CPU and an AMD
Radeon Pro 460 GPU using \texttt{futhark-bench}, which was configured
to report the average runtime of five different runs. Pricing of the
Vanilla option is based on 8388608 individual Monte Carlo simulation
paths, whereas pricing of the two other contracts is based on 1048576
individual paths. The Fut-C column shows timings for executables
generated using \texttt{futhark-c}, the CPU sequential-code compiler
for Futhark. The Fut-OpCL and the Fut-OpCL-Cut columns show timings
for executables generated with \texttt{futhark-opencl} and with the
Fut-OpCL-Cut column providing timings for the case where the $\cutPayoff{}$
functionality allows for pricing of the contract at different times
during the contract's lifetime.
\begin{table}
  \caption{Price timings in milliseconds. The measurements show the time it takes to price three different financial contracts using the FinPar Monte Carlo based generic pricing. The Fut-C column specifies sequential performance and the Fut-OpCL and the Fut-OpCL-Cut specify parallel performance with the Fut-OpCL-Cut column showing timings with the $\cutPayoff{}$ functionality enabled.   \label{timings.tab}}
  \vspace{-1em}
  \begin{tabular}{l|rrr} & Fut-C
                  & Fut-OpCL & Fut-OpCL-Cut \\ \hline
    Vanilla option & 6,779.4ms & 21.4ms & 21.7ms \\
    Barrier option & 1,521.7ms & 54.7ms & 55.1ms  \\
    Double option & 983.0ms & 12.4ms & 12.3ms
  \end{tabular}
\end{table}
The experiments show that for the vanilla European
option, a speedup of roughly 310 was achieved comparing the Futhark
program compiled into C (and further into x86 machine code) with a
version compiled into OpenCL using Futhark's OpenCL backend.

There are a number of observations to draw from the benchmark
results. First, notice that the speedup obtained from using the
commodity GPU instead of the laptop CPU ranges from a factor of 27 to
a factor of 317.\footnote{The multi-underlying nature of the
  non-vanilla contracts results in smaller speedups relative to the
  vanilla case due to the complexity and the sequential dependencies
  involved in dealing with correlations between underlyings.}
Second, notice that the introduction of the $\cutPayoff{}$ function in
the Fut-OpCL-Cut column has neglectable impact on performance. We can
therefore conclude that, at least for the contracts represented by the
three examples, the template feature makes it possible to avoid
recompilation of pricing code and that the generalisation can have a
dramatical positive effect on the performance of risk calculations,
each of which often consists of thousands of pricing tasks.

\section{Related Work}
There is a large body of work related to using domain specific
languages for specifying and managing financial contracts
\cite{SPJ2000,MLFi,SPJ2003,SimCorpXpress,hvitved10flacos,hvitved11jlap,BahrBertholdElsman}
and for specifying financial contract payoff expressions
\cite{Frankau09JFP}. Only parts of this work investigate the
certification aspects of the devised solutions
\cite{BahrBertholdElsman}. Compared to the previous work, the present
work considers how declarative certified contracts can be compiled
into generic payoff functions for efficient use in a practical pricing
framework.

Another line of related work investigates the possibility of
implementing financial contracts on distributed ledgers such as
blockchains \cite{Egelund-Muller2017}. Included in this work is work on establishing a
certified foundation for executing programs (also called smart
contracts) on such architectures \cite{O'Connor:2017:SNL:3139337.3139340}.

Finally, there is a large body of related work on developing
techniques for certifying implementations of programming languages,
including the seminal work on CompCert \cite{2006-Leroy-compcert}, a
fully certified compiler for the C programming language and the
verified LLVM project \cite{Zhao:2012:FLI:2103656.2103709}, which aims
at providing a pluggable toolkit for composing certified LLVM
\cite{Lattner:2004:LCF:977395.977673} compiler phases.

\section{Conclusion}
This work extends the certified contract management system of
\cite{BahrBertholdElsman} with template expressions, which allows for
drastic performance improvements and reusability in terms of the
concept of instruments (i.e., contract templates). We consider a
practical application of the declarative contract specifications in
the context of contract valuation (i.e., pricing). For the purposes of
interacting with pricing engines, we introduce a language for payoff
expressions (the payoff intermediate language). We have developed a
formalisation of the payoff intermediate language and a certified
compilation procedure in Coq. Our approach uses an extrinsic encoding,
which allows us to make use of Coq's code extraction feature for
obtaining a correct implementation of the compiler function that
translates expressions in CL to payoff expressions. We have introduced
a parameterisation technique for payoff expressions allowing for
capturing contract development over time. The developed technique is
consistent with the notion of contract reduction from
\cite{BahrBertholdElsman}.

A number of important properties, including soundness of the
translation from CL to the payoff language have been proved in Coq. We
have exemplified how the payoff intermediate language can be used to
generate code in a target language by mapping payoff expressions to a
subset of Haskell and Futhark. We have conducted performance
measurements with the generated Futhark code in the context of an
efficient parallel pricing engine and shown that, for three types of
contracts included in the experiment, the template scheme does not
significantly influence performance. On the contrary, the template
scheme allows for avoiding recompilation caused by changes to an
instrument's parameters and by simplification of a contract due to the
passage of time.

There are number of possibilities for future work.
First, some work is needed for the payoff intermediate language to
support the expression-level accumulation functionality from
\cite{BahrBertholdElsman}. As part of a solution, one may consider
generalising the somewhat ad-hoc $\<loopif>$ construct and, instead,
provide a more general language construct for iteration, which could
involve compiling $\<ifWithin>$ to a combination of iteration and
conditions (resulting in simpler target code generation).

Second, the representation of traces as functions $\nats ->
\type{Trans}$ is equivalent to infinite streams of transfers. It would
be interesting to explore this idea of using streams further, since
observable values also can be naturally represented as streams.

Finally, a possibility for future work is to formalise further the
infrastructure for working with external environment
representations. For instance, the reindexing scheme used in our
Futhark backend for accessing the external environment is currently
considered trusted code in the same way as the Futhark pretty
printing.


\begin{acks}                            
This research has been partially supported by the Danish Strategic
Research Council, Program Committee for Strategic Growth Technologies,
for the research center ``HIPERFIT: Functional High Performance
Computing for Financial Information Technology''
(\url{http://hiperfit.dk}) under contract number 10-092299
and by the CoqHoTT ERC Grant 637339. Any
opinions, findings, and conclusions or recommendations expressed in
this material are those of the authors and do not necessarily reflect
the views of the Danish Strategic Research Council.

\end{acks}

\bibliographystyle{plain}
\bibliography{paper}
\end{document}